\documentclass[runningheads]{llncs}

\usepackage[T1]{fontenc}
\usepackage{siunitx}
\usepackage{graphicx}
\usepackage{amsbsy}
\usepackage{color}
\usepackage{url}
\usepackage{ragged2e}

\begin{document}


\title{Robust Segmentation of Brain MRI in the Wild with Hierarchical CNNs and no Retraining}

\titlerunning{Robust Segmentation of Brain MRI in the Wild}

\author{Benjamin Billot \inst{1} \thanks{Corresponding author: benjamin.billot.18@ucl.ac.uk} \and
Colin Magdamo \inst{2} \and
Steven E. Arnold \inst{2} \and
Sudeshna Das \inst{2} \and
Juan Eugenio Iglesias\inst{1,3,4}}
\authorrunning{Billot, Magdamo, Arnold, Das and Iglesias}
\institute{Centre for Medical Image Computing, University College London, United Kingdom
\and Department of Neurology, Massachusetts General Hospital, USA
\and Martinos Center for Biomedical Imaging, Massachusetts General Hospital and Harvard Medical School, USA
\and Computer Science and Artificial Intelligence Laboratory, Massachusetts Institute of Technology, USA}

\maketitle

\begin{abstract}

Retrospective analysis of brain MRI scans acquired in the clinic has the potential to enable neuroimaging studies with sample sizes much larger than those found in research datasets. However, analysing such clinical images ``in the wild'' is challenging, since subjects are scanned with highly variable protocols (MR contrast, resolution, orientation, etc.). Nevertheless, recent advances in convolutional neural networks (CNNs) and domain randomisation for image segmentation, best represented by the publicly available method \textit{SynthSeg}, may enable morphometry of clinical MRI at scale. In this work, we first evaluate \textit{SynthSeg} on an uncurated, heterogeneous dataset of more than 10,000 scans acquired at Massachusetts General Hospital. We show that \textit{SynthSeg} is generally robust, but frequently falters on scans with low signal-to-noise ratio or poor tissue contrast. Next, we propose \textit{SynthSeg$^{+}$}, a novel method that greatly mitigates these problems using a hierarchy of conditional segmentation and denoising CNNs. We show that this method is considerably more robust than \textit{SynthSeg}, while also outperforming cascaded networks and state-of-the-art segmentation denoising methods. Finally, we apply our approach to a proof-of-concept volumetric study of ageing, where it closely replicates atrophy patterns observed in research studies conducted on high-quality, \SI{1}{\milli\meter}, T1-weighted scans. The code and trained model are publicly available at \url{https://github.com/BBillot/SynthSeg}.


\end{abstract}

\section{Introduction}

Neuroimaging with MRI is of paramount importance in the understanding of the morphology and connectivity of the human brain. \textit{In vivo} MR imaging has been widely adopted in research, where studies rely most often on prospective datasets of high-quality brain scans. Meanwhile, clinical MRI datasets remain largely unexplored in neuroimaging studies, despite their much higher abundance (e.g., 10 million brain scans were acquired in the US in 2019~\cite{oren_curbing_2019}). Analysing such datasets is highly desirable, since it would enable sample sizes in the millions, which is much higher than the current largest research studies, which include tens of thousands subjects (e.g., ENIGMA~\cite{hibar_common_2015} or UK BioBank~\cite{alfaro-almagro_image_2018}).

The use of clinical data in neuroimaging studies has been mainly hindered by the high variability in acquisition protocols. As opposed to high resolution (HR) scans used in research, physicians usually prefer low resolution (LR) acquisitions with fewer slices (for faster inspection), which span a large range of orientations, slice spacings and slice thicknesses. Moreover, clinical scans employ numerous MRI contrasts to highlight different tissue properties. 

Overall, no segmentation method can robustly adapt to such variability. Manual labelling is the gold standard in segmentation techniques, but it remains too tedious for large-scale clinical applications. An alternative would be to only consider subjects with high-quality acquisitions (e.g., \SI{1}{\milli\meter} T1 scans), as these can be easily analysed with neuroimaging softwares~\cite{ashburner_unified_2005,fischl_freesurfer_2012}. However, this would enormously decrease the effective sample size of clinical datasets, where such scans are seldom available. Hence, there is a clear need for a robust automated segmentation tool that can adapt to clinical scans of any MRI contrast and~resolution.

Contrast-invariance has traditionally been achieved via Bayesian segmentation strategies with unsupervised likelihood model~\cite{puonti_fast_2016}. Unfortunately, these methods are highly sensitive to partial volume effects (PV) caused by changes in resolution~\cite{choi_partial_1991}. This problem can partly be mitigated by directly modelling PV within the Bayesian framework~\cite{van_leemput_unifying_2003}. However, this strategy quickly becomes intractable for scans with decreasing resolutions and increasing number of labels, thus limiting its application to large clinical datasets.

Recent automated segmentation methods rely on supervised convolutional neural networks (CNNs)~\cite{kamnitsas_efficient_2017,milletari_v-net_2016,ronneberger_u-net_2015}. While CNNs obtain fast and accurate results on their training domain, they are fragile to changes in resolution~\cite{ghafoorian_transfer_2017,orbes-arteaga_multi-domain_2019} and MRI contrast~\cite{akkus_deep_2017,jog_psacnn_2019}, even within the same MRI modality~\cite{karani_test-time_2021,kushibar_supervised_2019}. Although data augmentation can improve robustness in intra-modality scenarios~\cite{zhang_generalizing_2020}, CNNs still need to be retrained for each new MRI contrast and resolution. This issue has sparked a vivid interest in domain adaptation schemes, where CNNs are trained to generalise to a specific target domain~\cite{chen_synergistic_2019,karani_test-time_2021}. However, these methods still need to be retrained for every new target resolution or MRI contrast, which makes them impractical to apply on highly heterogeneous clinical data.

Very recently, a publicly available method named \textit{SynthSeg}~\cite{billot_synthseg_2021} has been proposed for out-of-the-box segmentation of brain scans of any contrast and resolution. \textit{SynthSeg} relies on a 3D UNet trained on synthetic scans generated with a domain randomisation approach~\cite{tobin_domain_2017}. While \textit{SynthSeg} yields excellent generalisation compared with previous techniques, it still lacks robustness when applied to clinical scans with low signal-to-noise (SNR) ratio or poor tissue contrast. 

Improvements in robustness have previously been tackled with hierarchical models, where a first CNN performs a simpler preliminary task (e.g., predicting an initial mask~\cite{roth_application_2018}, or pre-segmenting at low resolution~\cite{isensee_nnu-net_2021}), and the results are refined by a second network trained for the target task. However, these methods often remain insufficient to capture high-order topological relations, which is a well-known problem for CNNs~\cite{nosrati_incorporating_2016}. A possible solution is to use conditional random fields for postprocessing~\cite{kamnitsas_efficient_2017}, but these often struggle to model relations between multiple labels at different scales. Recent methods now seek to improve semantic correctness either by aligning predictions and ground truths in latent space during training~\cite{oktay_anatomically_2018}, or by using denoising CNNs~\cite{larrazabal_post-dae_2020}. Although these methods have shown promising results in relatively simple cases (i.e., 2D images with few labels), they are yet to be demonstrated in more complex setups.

In this work, we present \textit{SynthSeg$^{+}$}, a novel architecture for robust segmentation of clinical MRI scans of any contrast and resolution without retraining. Specifically, we build on the domain randomisation strategy introduced by \textit{SynthSeg}, and propose a hierarchy of conditional segmentation and denoising CNNs for improved robustness and semantic correctness. We evaluate this method on more than 10,000 highly heterogeneous clinical scans, directly taken from the picture archiving communication system (PACS) of Massachusetts General Hospital (MGH). \textit{SynthSeg$^{+}$} yields considerably enhanced robustness compared to \textit{SynthSeg}, while also outperforming cascaded networks and state-of-the-art denoising methods.

\section{Methods}

\subsection{Hierarchical conditional architecture}

\begin{figure}[t]
\includegraphics[width=\textwidth]{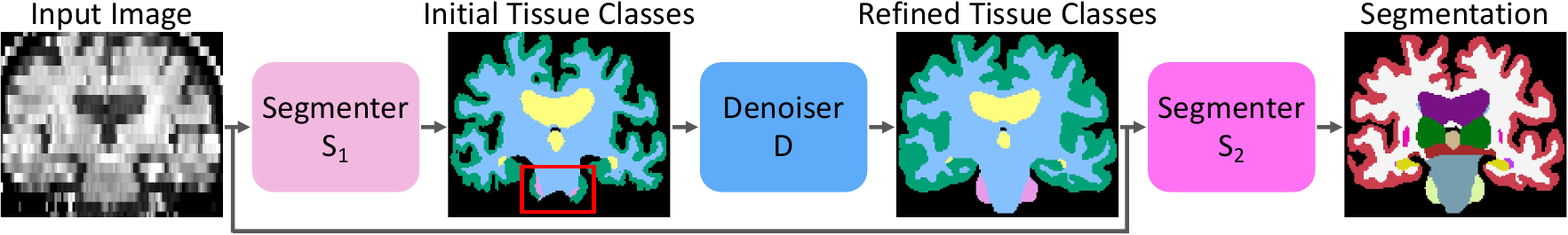}
\caption{Overview of the proposed architecture. A first network $S_1$ outputs initial segmentations of four tissue classes. Robustness is then improved  by refining these with a denoiser $D$ (e.g., red box). Final segmentations are obtained with a second segmenter $S_2$, which takes as inputs the image and the robust estimates of the four tissue classes.}
\label{fig:archi}
\end{figure}

We propose an architecture that relies on three hierarchical CNN modules (Fig.~\ref{fig:archi}). This design aims at efficiently subdividing the target segmentation task into intermediate steps that are easier to perform, and thus less prone to errors. For this purpose, a first network $S_1$ is trained to produce coarse initial segmentations of the input images. More precisely, these initial segmentations only contain four labels that group brain regions into classes of similar tissue types and intensities (cerebral white matter, cerebral grey matter, cerebrospinal fluid, and cerebellum). These classes are easier to discriminate than individual regions.

The output of $S_1$ is then fed to a denoising network $D$~\cite{larrazabal_post-dae_2020} in order to increase the robustness of the initial segmentations. By modelling high-level relations between tissue types, $D$ seeks to correct potential semantic inconsistencies introduced by $S_1$ (e.g, cerebral grey matter in the cerebellum). Moreover, it also enables recovery from large mistakes in the initial tissue classes, which sometimes occur for scans with low SNR, poor contrast, or very low resolution.

Final segmentations are obtained with a second segmenter $S_2$, which takes as input the image and the corrected tissue classes given by $D$, thus combining the robustness of $D$ with the accuracy of $S_2$. In practice, $S_2$ learns to subdivide the initial tissue predictions into the target labels, as well as to refine the boundaries given by $D$, which are often excessively smooth (e.g., the cortex in Fig.~\ref{fig:archi}).

\subsection{Training scheme for the segmentation modules}
\label{sec:training_s}

\begin{figure}[t]
\centering
\includegraphics[width=\textwidth]{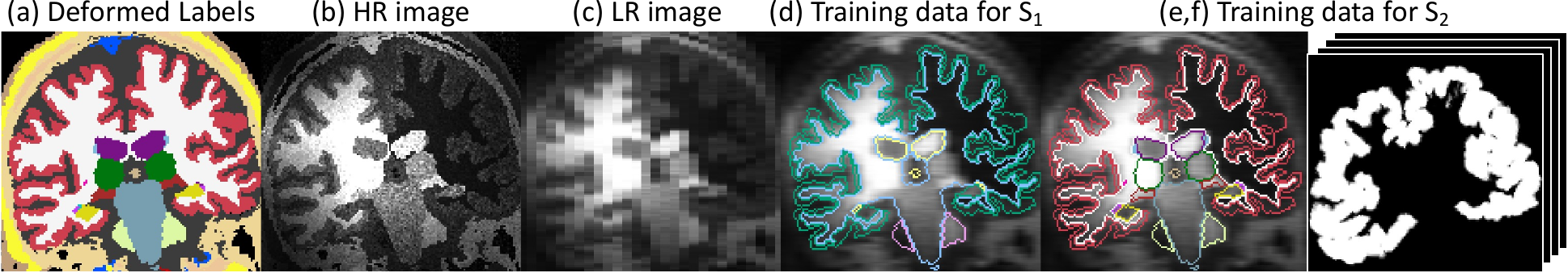}
\caption{
Generative model to train $S_1$ and $S_2$. A label map is deformed~(a) and used to generate an HR scan~(b), from which we simulate a LR scan of random resolution~(c). $S_1$ is trained to produce HR segmentations of the four coarse tissue classes from the LR image, which is upsampled to HR space for convenience~(d). $S_2$ is trained to estimate all target labels, using both the image and soft tissue segmentations~(e,f). During training, we corrupt these tissue maps to model the errors made by the denoiser $D$.
}
\label{fig:generation}
\end{figure}

$S_1$ and $S_2$ are trained separately with a domain randomisation strategy~\cite{tobin_domain_2017}. 
Specifically, we use synthetic data sampled on the fly from a generative model that only needs a set of label maps as input, and whose parameters (contrast, resolution, artefacts, etc.) are drawn from uninformative priors (Supplementary Table~\ref{tab:priors}). As a result, $S_1$ and $S_2$ are exposed to vastly varying examples, which forces them to learn contrast- and resolution-agnostic features. Training image-target pairs for $S_1$ are generated with a procedure similar to \textit{SynthSeg}~\cite{billot_synthseg_2021}:

\textit{(a)}
We draw a label map from a set of 3D segmentations with $N$ labels. We assume that these are defined on a grid of $J$ voxels at high resolution $r_{HR}$ (here \SI{1}{\milli\meter} isotropic). We then spatially augment the segmentation with a nonlinear transform (a smooth stationary velocity field~\cite{arsigny_log-euclidean_2006}) as well as three rotations, scalings, shearings, and translations. We call this augmented map $L$ (Fig.~\ref{fig:generation}a).

\textit{(b)} 
We obtain an image $G=\{G_j\}_{j=1}^J$ by sampling a Gaussian Mixture Model (GMM) conditioned on $L$. All means and variances $\{\mu_n, \sigma^2_n\}_{n=1}^N$ are sampled from uniform priors to obtain a different random contrast at each minibatch~\cite{billot_learning_2020}. We also corrupt $G$ by a random bias field $B$, sampled in logarithmic domain:
\begin{equation}
    p(G| L, \{\mu_n, \sigma^2_n\}_{n=1}^N) = 
    \prod_{j=1}^J   \frac{B_j}{\sqrt{2\pi \sigma^2_{L_j}}} \exp [ -\frac{1}{2  \sigma^2_{L_j}} ( G_j B_j - \mu_{L_j} )^2]
\end{equation}
An image $I_{HR}$ is then formed by normalising the intensities of $G$ in $[0, 1]$, and nonlinearly augmenting them with a random voxel-wise exponentiation (Fig.~\ref{fig:generation}b).

\textit{(c)}
We then simulate LR scans with PV. This is achieved by blurring $I_{HR}$ with a Gaussian kernel $K$ of random standard deviation (to model slice thickness), and subsampling it to a random low resolution $r_{sp}$ (to simulate slice spacing). Finally, we form an image $I_{LR}$ by modelling the scanner noise with an additive field $\mathcal{E}$, sampled from a zero-mean Gaussian of random variance~(Fig.~\ref{fig:generation}c):
\begin{equation}
    I_{LR} = Resample(I_{HR} \ast K, r_{sp}) + \mathcal{E}. 
\end{equation}

\textit{(d)} 
The final training image $I$ is obtained by resampling $I_{LR}$ back to $r_{HR}$, while the target segmentation for $S_1$ is built by grouping the cerebral regions of $L$ into the four tissue classes (Fig.~\ref{fig:generation}d). We emphasise that test scans will also be resampled to $r_{HR}$, since this enables us to segment on the target HR grid.

Training data for $S_2$ are sampled from the same model, but with two differences. First, the ground truths now contain all target labels (Fig.~\ref{fig:generation}e). Second, we still build tissue segmentations as in \textit{(d)} (since these are needed as inputs to $S_2$), but we now represent them as soft probability maps, which we randomly dilate/erode and spatially deform to model $D$ imperfections at test-time (Fig.~\ref{fig:generation}f).

\subsection{Training scheme for the denoising module}
\label{sec:training_d}

\begin{figure}[t]
\includegraphics[width=\textwidth]{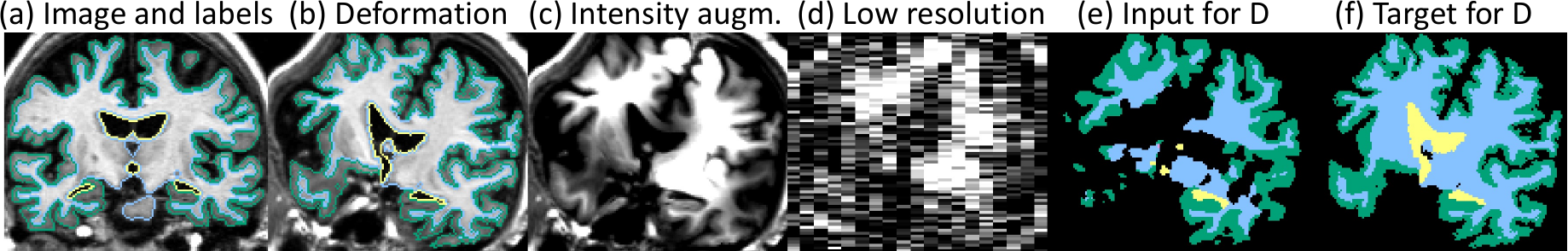}
\caption{Degradation model to train $D$. (a) A real image and its labels are spatially~deformed (b, f). The image is degraded (c,d) and fed to $S_1$ to obtain the input for $D$~(e).}
\label{fig:degradation}
\end{figure}

Recent denoising methods are mostly based on supervised CNNs trained to recover ground truth segmentations from artificially corrupted versions of the same maps~\cite{karani_test-time_2021,khan_deep_2021,larrazabal_post-dae_2020}. However, the employed corruption strategies are often handcrafted (random erosion and dilation, swapping of labels, etc.), and thus do not accurately capture errors made by the segmentation method to correct. 

Instead, we propose to employ examples representative of $S_1$ errors, obtained by degrading real images, and feeding them to the trained $S_1$. $D$ is then trained to map the outputs of $S_1$ back to their ground truths. Images are degraded on the fly with the same steps as Section~\ref{sec:training_s} (except for the GMM, since we now use real images): spatial deformation, bias field, voxel-wise exponentiation, simulation of low resolution, and noise injection (Fig.~\ref{fig:degradation}). In practice, these corruptions use considerably wider prior parameter distributions than in Section~\ref{sec:training_s} (Table~\ref{tab:priors}), in order to ensure a high probability of erroneous segmentations from $S_1$.

\subsection{Implementation details}

All modules are trained separately using the average soft Dice as loss function~\cite{milletari_v-net_2016}. The segmentation modules $S_1$ and $S_2$ use the same 3D UNet architecture as in~\cite{billot_synthseg_2021}. Briefly, it comprises 5 levels of 2 convolution layers. All layers use 3$\times$3$\times$3 kernels and ELU activation~\cite{clevert_fast_2016}, except for the last layer, which employs a softmax. The first layer has 24 feature maps; this number is doubled after each max-pooling, and halved after each upsampling. Meanwhile, the denoiser $D$ uses a lighter structure chosen with a validation set: one convolution per layer, with a constant number of 16 features. Importantly, we delete the skip connections between the top two levels, to reach a compromise between UNets, where high-level skip connections propagate potential errors in the input segmentations to correct; and auto-encoders, with excessive bottleneck-induced smoothness.

\section{Experiments and Results}

\subsection{Brain MRI Datasets}

The training dataset consists of 1020 \SI{1}{\milli\meter} T1 brain scans (500 from ADNI~\cite{jack_alzheimers_2008}, 500 from HCP~\cite{van_essen_human_2012}, 20 from a private dataset~\cite{fischl_whole_2002}), which are available with a combination of manual and automated labels for 44 regions (Table~\ref{tab:regions}). We note that $S_1$ and $S_2$ are only trained with the label maps, and that using subjects from varied populations enables us to increase robustness to morphological variability.

We evaluate \textit{\textit{SynthSeg}$^{+}$} on 10,520 uncurated scans from the PACS of MGH. These are obtained from 1,047 MRI sessions of distinct subjects (Fig. S3), using a huge range of resolutions and contrasts (T1, T2, FLAIR, B0 channels from diffusion, among others). Among all sessions, 62 include T1 scans at maximum \SI{1.3}{\milli\meter} resolution, for which we obtain label maps by running FreeSurfer~\cite{fischl_freesurfer_2012}. Next, we propagate the resulting labels with rigid registration to all the scans of the corresponding sessions, which provides us with silver standard segmentations for 520 scans. Finally, these are split between validation (20), and testing (500), while the other 10,000 scans are used for indirect evaluation.

\subsection{Competing Methods}

We evaluate \textit{\textit{SynthSeg}$^{+}$} (i.e., $S_1 + D + S_2$) against four approaches.

\noindent\textbf{\textit{SynthSeg}~\cite{billot_synthseg_2021}}: We use the publicly available model for testing.

\noindent\textbf{Cascaded networks~\cite{roth_application_2018} ($\boldsymbol{S_1 + S_2}$)}: we ablate the denoiser $D$ to obtain an architecture that is representative of classical cascaded networks.

\noindent\textbf{Denoiser~\cite{larrazabal_post-dae_2020} (\textit{SynthSeg} $\boldsymbol{+ D}$)}: A state-of-the-art method for denoising by postprocessing, where a denoiser ($D$) is appended to the method to correct (\textit{SynthSeg}). Here, $D$ is trained as in Section~\ref{sec:training_d} to correct \textit{all} target labels.

\noindent\textbf{Cascaded networks with appended denoiser ($\boldsymbol{S_1 + S_2 + D}$)}: A combination of the cascaded architecture with the denoising network $D$.

All networks are trained for 300,000 steps (7 days on a Nvidia RTX6000) with the Adam optimiser~\cite{kingma_adam_2017}. Based on~\cite{billot_synthseg_2021}, we use Keras~\cite{chollet_keras_2015} and Tensorflow~\cite{abadi_tensorflow_2016}. Inference takes between 8 and 12 seconds for all methods on the same GPU.

\subsection{Quantitative analysis}

\begin{figure}[t]
\includegraphics[width=\textwidth]{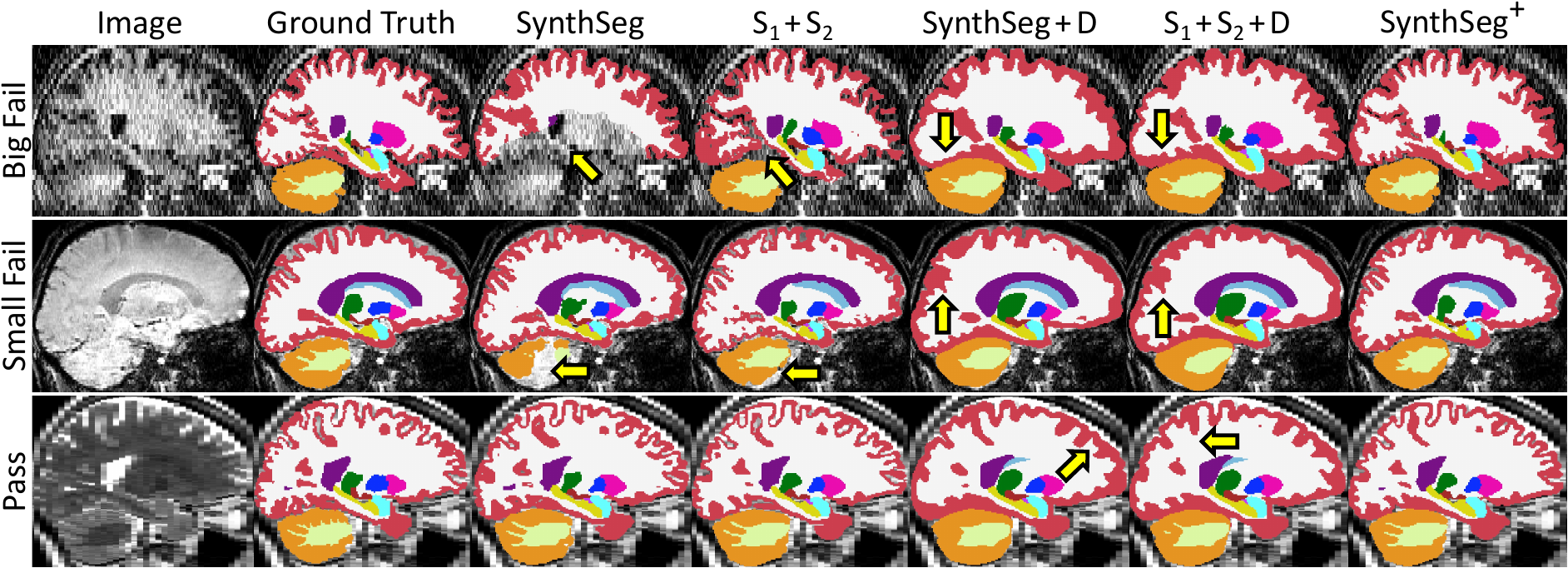}
\caption{Segmentations obtained by all methods for scans where \textit{SynthSeg} shows large (``Big Fail''),  mild (``Small Fail''), or no errors (``Pass''). Arrows indicate major mistakes. \textit{SynthSeg$^{+}$} yields outstanding results given the low SNR, poor tissue contrast, or low resolution of the inputs. Note that appending $D$ considerably smooths segmentations.}
\vspace{0.1cm}
\label{fig:segmentations}
\end{figure}

\begin{figure}[t]
\begin{center}
\includegraphics[width=\textwidth]{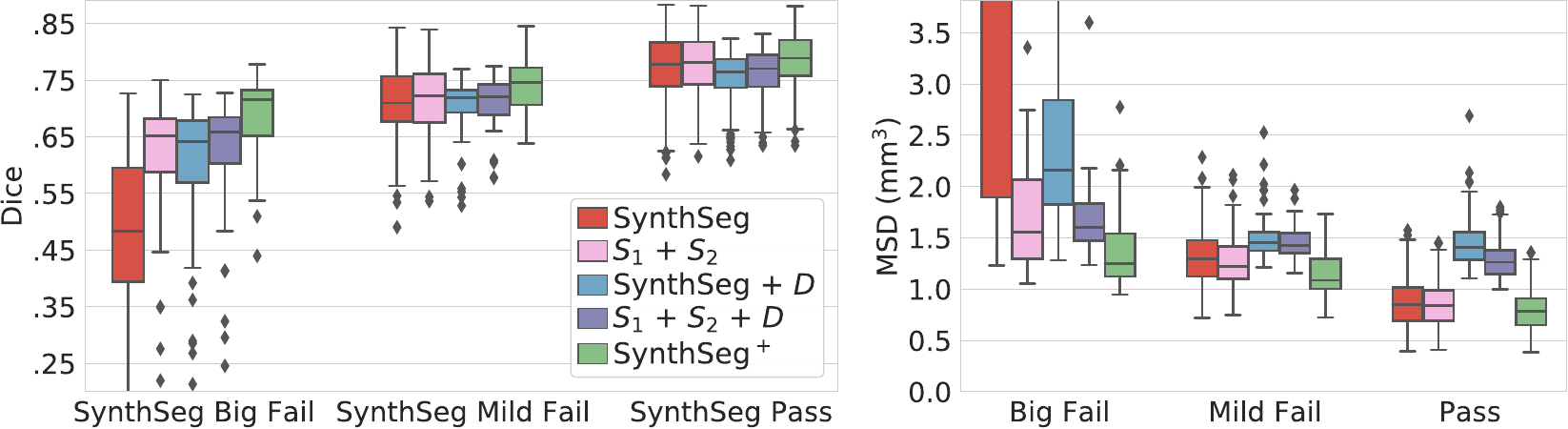}
\caption{Dice scores and mean surface distances (MSD) for 500 heterogeneous clinical scans. Results are presented based on the outcome of a QC analysis performed on the segmentations of \textit{SynthSeg}: Big Fails (82 cases), Mild Fails (103), and Passes (315).}
\vspace{-0.6cm}
\label{fig:dice}
\end{center}
\end{figure}

First, we evaluate all methods on the 500 scans with ground truth. For visualisation purposes, we subdivide these scans into three classes based on the outcome of a visual quality control (QC) performed on the segmentations of \textit{SynthSeg} (Fig.~\ref{fig:segmentations}, third column): large errors (``Big fails'', 82 cases), mild errors (``Small Fails'', 103 cases) and good segmentations (``Passes'', 315 cases).

Figure~\ref{fig:dice} reveals that decomposing the target segmentation task into easier steps considerably improves robustness~\cite{roth_application_2018}. Indeed, the cascaded networks $S_1 + S_2$ outperforms \textit{SynthSeg} by 16.5 Dice points on the worst scans (i.e., Big Fails), while presenting much less outliers. Moreover, this strategy slightly increases accuracy, and leads to improvements by up to 2 points for small fails.

In comparison, using a denoiser for postprocessing also improves robustness, but leads to a non-negligible loss in accuracy for Mild Fails and Passes, leading to an increase of at least \SI{0.25}{\milli\meter} in mean surface distance (MSD) compared with \textit{SynthSeg}. This is due to the fact that $D$ cannot accurately model convoluted boundaries and returns very smoothed segmentations (e.g., the cortex in Fig.~\ref{fig:segmentations}). Moreover, since $D$ does not have access to the input scan, its segmentations may deviate from the original anatomy, which may introduce biases when analysing populations with morphologies different from the training set.

Remarkably, \textit{SynthSeg$^{+}$} yields an outstanding robustness (see Fig. S4 for examples of failures) and obtains the best scores in all three categories, with an outstanding improvement of 23.5 Dice points over \textit{SynthSeg} for Big Fails. In comparison with cascaded networks, inserting a denoiser $D$ between $S_1$ and $S_2$ enables us to obtain robust tissue segmentations, which consistently improves scores by 2 to 5 Dice points. Finally, integrating $D$ within our framework (rather than using it for postprocessing) enables \textit{SynthSeg$^{+}$} to exploit both images and prior information when predicting final segmentations. As a result, our approach is more accurate than the two methods using $D$ for postprocessing (Fig.~\ref{fig:segmentations}), and outperforms them by at least \SI{0.35}{\milli\meter} in MSD and 3 Dice points.

\subsection{Volumetric study}
\label{sec:vol_study}

\begin{figure}[t]
\centering
\includegraphics[width=\textwidth]{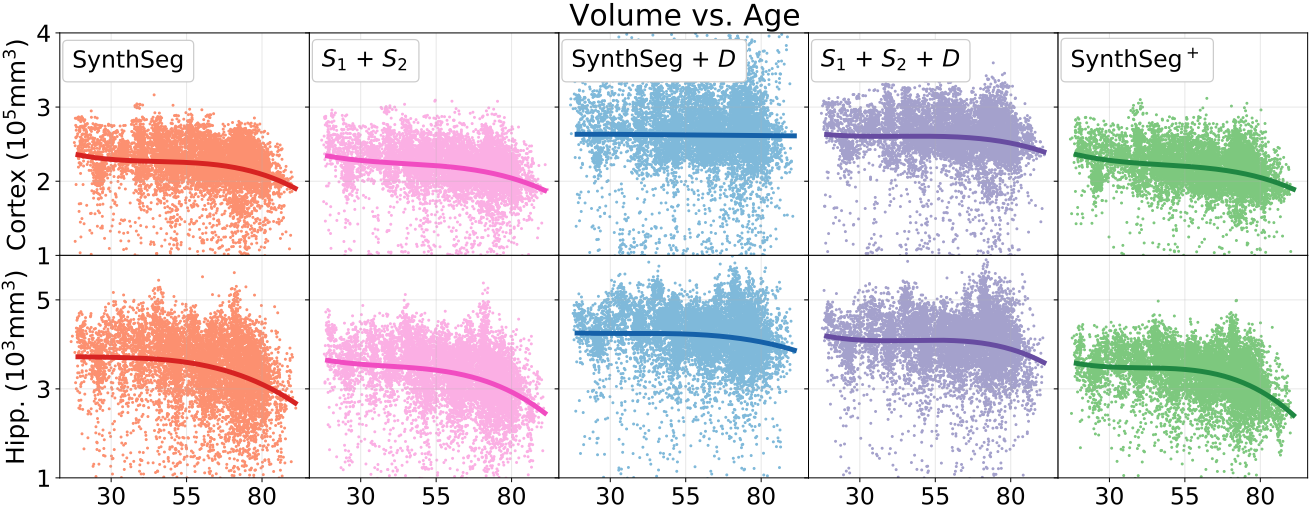}
\caption{Trajectories of cortical and hippocampal volumes with age (10,000 scans).}
\label{fig:volumes}
\end{figure}

We now conduct a proof-of-concept volumetric study on the held-out 10,000 scans. Specifically, we analyse age-related atrophy using the volumes estimated by all methods. Our ageing model includes: B-splines with 10 equally spaced knots and soft constraints for monotonicity, linear terms for slice spacing in each acquisition direction, and a bias for gender. We then fit this model numerically by minimising the sum of squares of the residuals with the L-BFGS-B method~\cite{byrd_limited_1995}.

Figure~\ref{fig:volumes} shows the scatter plot and B-spline fits for the cerebral cortex and hippocampus (see Fig. S5 for the other regions). It reveals that the two methods using $D$ for postprocessing detect no or small atrophy, and over-estimate the volumes relatively to the other methods (due to the smoothing effect). Remarkably, \textit{SynthSeg$^{+}$} yields atrophy curves very close to those obtained in recent studies conducted on scans of much higher quality (i.e., \SI{1}{\milli\meter} T1 scans)~\cite{coupe_towards_2017,dima_subcortical_2022}. While \textit{SynthSeg} and $S_1 + S_2$ yield similar average trajectories, \textit{SynthSeg$^{+}$} produces far fewer outliers (especially at LR, see Fig. S6), which suggests that it can be used to investigate other population effects with much higher statistical power.

\section{Conclusion}

We have presented \textit{SynthSeg$^{+}$}, a novel hierarchical architecture that enables large-scale robust segmentation of brain MRI scans in the wild, without retraining. Our method shows considerably improved robustness relatively to \textit{SynthSeg}, while outperforming cascaded CNNs and state-of-the-art denoising networks. We demonstrate \textit{SynthSeg$^{+}$} in a study of ageing using 10,000 highly heterogeneous clinical scans, where it accurately replicates atrophy patterns observed on research data of much higher quality. By releasing the trained model, we aim at greatly facilitating the adoption of neuroimaging studies in the clinic, which has the potential to highly improve our understanding of neurological disorders.

\subsubsection{Acknowledgement} This work is supported by the European Research Council (ERC Starting Grant 677697), the EPSRC-funded UCL Centre for Doctoral Training in Medical Imaging (EP/L016478/1), the Department of Health's NIHR-funded Biomedical Research Centre at UCLH, Alzheimer’s Research UK (ARUK-IRG2019A-003), and the NIH (1R01AG070988, 1RF1MH123195).

\bibliographystyle{splncs04}


\newpage
\renewcommand{\thetable}{S\arabic{table}}
\renewcommand{\thefigure}{S\arabic{figure}}
\makeatother
\makeatletter

\title{Robust Segmentation of Brain MRI in the Wild \\ Billot, Magdamo, Arnold, Das and Iglesias \\ Supplementary Materials }
\titlerunning{Robust Segmentation of Brain MRI with a Single Hierarchical CNN}
\author{}
\institute{}
\maketitle 

\vspace{-1.3cm}
\begin{table}
\setlength\tabcolsep{7.5pt}
\caption{Ranges of the uniform distributions used for the generative model and the degradation model. The GMM assumes intensities in [0, 255]. The nonlinear transform and bias field are modelled by sampling a small tensor from a zero-mean Gaussian (respectively of variance $\sigma_v^2$ and $\sigma_B^2$), and upsampling it to $r_{HR}$. $\gamma$ is the value of the voxel-wise exponentiation. $\sigma_{th}$ and $\sigma_\mathcal{E}$ are the standard deviations of $K$ and $\mathcal{E}$.}
\vspace{-0.25cm}
\centering
\label{tab:priors}
\begin{tabular}{|l|c|c|}
\hline
Parameters & Generative model prior & Degradation model prior \\
\hline
Rotation $(^\circ)$ & [-15, 15] & [-25, 25] \\
Scaling & [0.85, 1.15] & [0.5, 1.5] \\
Shearing & [-0.012, 0.012] & [0.02, 0.02] \\
Translation & [-20,20] & [-50, 50] \\
Nonlinear transform $\sigma_v^2$ & [0, 1.5] & [0, 4] \\
GMM means $\mu_k$ & [0, 255] & - \\
GMM variances $\sigma_k^2$ & [0, 6] & - \\
Bias field $\sigma_B^2$ & [0, 0.25] & [0, 3] \\
Intensity corruption $\gamma$ & [0.9, 1.1] & [0.3, 3] \\
Slice thickness $\sigma_{th}$ & [0.5, 5] & [0.5, 8] \\
Slice spacing $r_{sp}$ & [1, 9] & [1, 12] \\
Noise injection $\sigma_\mathcal{E}$ & [0, 10] & [0, 50] \\
\hline
\end{tabular}
\end{table}

\vspace{-0.7cm}
\begin{table}
\setlength\tabcolsep{4pt}
\caption{List of the structures present in the training maps. Labels with different contralateral values are marked with $^{\text{R/L}}$, whereas predicted regions are noted with *.}
\vspace{-0.25cm}
\centering
\label{tab:regions}
\begin{tabular}{|c|c|c|c|}
\hline
Background*  & Thalamus$^{\text{R/L}}$* &  Hippocampus$^{\text{R/L}}$*  &  Artery\\
Cerebral white matter$^{\text{R/L}}$*  & Caudate$^{\text{R/L}}$*  &  Amygdala$^{\text{R/L}}$*  &  Optic chiasm \\
Cerebral cortex$^{\text{R/L}}$* &  Putamen$^{\text{R/L}}$*  &  Accumbens area$^{\text{R/L}}$*  &  Soft tissues \\
Lateral ventricle$^{\text{R/L}}$*  &  Pallidum$^{\text{R/L}}$*  &  Ventral DC$^{\text{R/L}}$*  &  Mucosa\\
Inferior Lateral Ventricle$^{\text{R/L}}$*  &  $3^{rd}$ ventricle*  &  Cerebral vessels$^{\text{R/L}}$  &  Skin\\
Cerebellar white matter$^{\text{R/L}}$*   &  $4^{th}$ ventricle*  &  Choroid plexus$^{\text{R/L}}$  &  Skull\\
Cerebellar grey matter$^{\text{R/L}}$*  &  Brainstem*   &  CSF & Eyes \\
\hline
\end{tabular}
\end{table}

\vspace{-0.4cm}
\begin{center}
\includegraphics[width=0.95\textwidth]{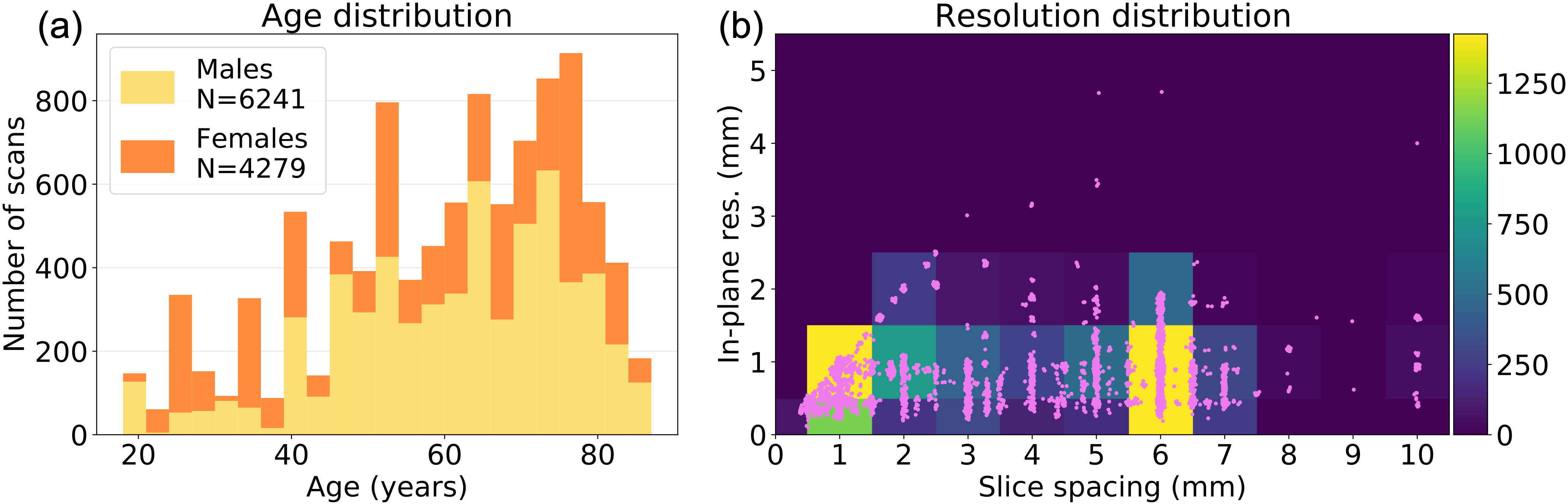} \\
\justifying
\noindent\textbf{Figure S3.} (a) Age/gender and (b) resolution distributions (individual scans in pink) for 10,520 scans acquired at MGH during 1,047 subject sessions.
\end{center}

\vspace{0.2cm}
\begin{center}
\justifying
\includegraphics[width=\textwidth]{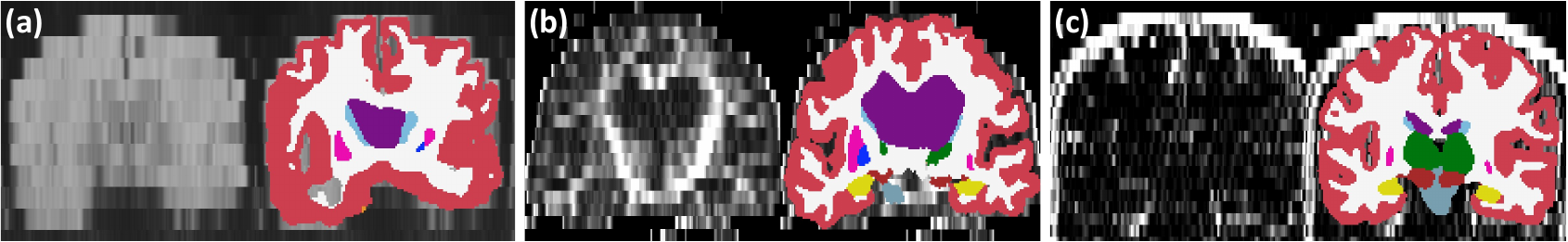} \\
\small \textbf{Figure S4.} The three scans for which \textit{SynthSeg$^{+}$} obtains the worst Dice scores in the quantitative analysis experiment. These scans are very challenging to segment, since they are all at very low resolution, and present (a) almost no tissue contrast, (b) very low signal-to-noise ratio, or (c) even both combined. Nevertheless, we see that \textit{SynthSeg$^{+}$} already produces a good segmentation in the third case.
\end{center}

\vspace{0.2cm}
\begin{center}
\justifying
\includegraphics[width=\textwidth]{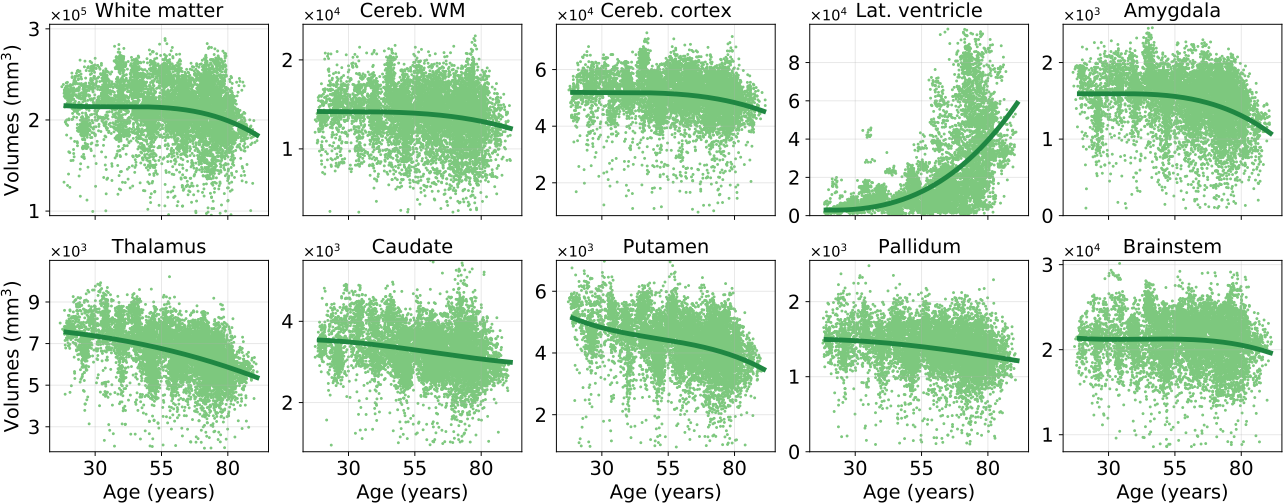} \\
\small \textbf{Figure S5.} Volume trajectories for representative regions obtained with \textit{SynthSeg$^{+}$} on the 10,000 scans of Section~\ref{sec:vol_study}. We emphasise that these curves are very similar to those obtained in recent research studies (Coup\'e et al., 2017; Dima et al., 2022), which employed much higher quality scans (\SI{1}{\milli\meter} isotropic MP-RAGE sequences).
\end{center}

\vspace{0.2cm}
\begin{center}
\justifying
\includegraphics[width=\textwidth]{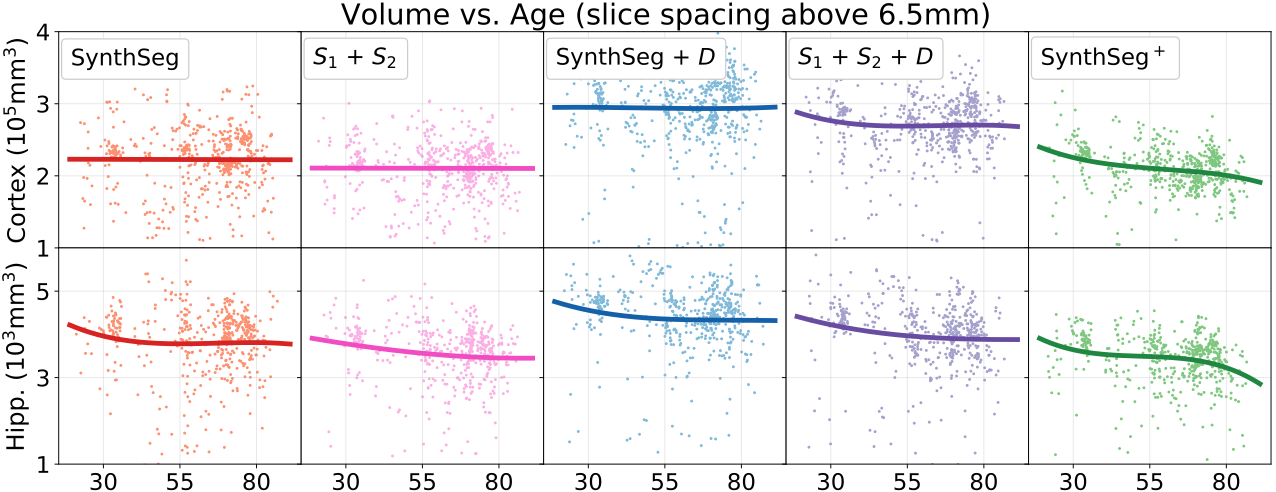} \\
\textbf{Figure S6.} Volume trajectories computed for scans with slice spacing above \SI{6.5}{\milli\meter} (N$=587$). The competing approaches yield very uniform volumes across the age span, and produce many outliers. In contrast, \textit{SynthSeg$^{+}$} is much more robust: it detects atrophy patterns very similar to those obtained on all available subjects, and presents much less outliers than the other methods.
\end{center}

\end{document}